\documentclass[journal=jctc,manuscript=article]{achemso}

\usepackage{amssymb}
\usepackage{amsmath}
\usepackage{mathtools}
\usepackage{graphicx}
\usepackage{float}
\usepackage{braket}
\usepackage[version=3]{mhchem} 
\usepackage[T1]{fontenc}       
\usepackage{caption}
\usepackage{lipsum}
\usepackage{subfig}
\usepackage{xcolor}
\usepackage{siunitx}

\newcommand*{\citen}[1]{%
  \begingroup
    \romannumeral-`\x 
    \setcitestyle{numbers}%
    \cite{#1}%
    \endgroup
}

\author{Wenjie Dou} \affiliation{Department of Chemistry, University
  of California Berkeley, Berkeley California 94720, USA}
\email{douw@berkeley.edu}

\author{Tyler Y. Takeshita} \email{tyler.takeshita@daimler.com}
\affiliation{Mercedes-Benz Research and Development North America,
  Sunnyvale, CA 94085}

\author{Ming Chen} \affiliation{Department of Chemistry, University of
  California Berkeley, Berkeley California 94720, USA}
\alsoaffiliation{Materials Sciences Devision, Lawrence Berkeley
  National Laboratory, Berkeley, California 94720, USA}
\email{mingchen.chem@berkeley.edu}

\author{Roi Baer} \email{roi.baer@huji.ac.il} \affiliation{Fritz Haber
  Research Center for Molecular Dynamics, Institute of Chemistry, The
  Hebrew University of Jerusalem, Jerusalem 9190401, Israel}

\author{Daniel Neuhauser} \email{dxn@chem.ucla.edu}
\affiliation{Department of Chemistry and Biochemistry, University of
  California, Los Angeles, California 90095, USA}

\author{Eran Rabani} \email{eran.rabani@berkeley.edu}
\affiliation{Department of Chemistry, University of California
  Berkeley, Berkeley California 94720, USA} \alsoaffiliation{Materials
  Sciences Devision, Lawrence Berkeley National Laboratory, Berkeley,
  California 94720, USA} \altaffiliation{The Raymond and Beverly
  Sackler Center of Computational Molecular and Materials Science, Tel
  Aviv University, Tel Aviv 69978, Israel}

\nopagebreak
\title{Stochastic Resolution of Identity for Real-Time Second-Order
  Green's Function: Ionization Potential and Quasi-particle Spectrum}
\nopagebreak
\abbreviations{}
\keywords{Resolution of identity, Stochastic orbitals}

\nopagebreak
\SectionNumbersOn
\nopagebreak

\begin{document}


\nopagebreak
\begin{abstract}
  We develop a stochastic resolution of identity approach to the
  real-time second-order Green's function (real-time sRI-GF2) theory,
  extending our recent work for imaginary-time Matsubara Green's
  function {\em J. Chem. Phys.} {\bf 151}, 044114 (2019)).  The
  approach provides a framework to obtain the quasi-particle spectra
  across a wide range of frequencies as well as predict ionization
  potentials and electron affinities.  To assess the accuracy of the
  real-time sRI-GF2, we study a series of molecules and compare our
  results to experiments and to a many-body perturbation approach
  based on the GW approximation, where we find that the real-time
  sRI-GF2 is as accurate as self-consistent GW.  The stochastic
  formulation reduces the formal scaling to $O(N_e^3)$, where $N_e$ is
  the number of electrons. This is illustrated for a chain of hydrogen
  dimers, where we observe a slightly lower than cubic scaling for
  systems containing up to $N_e \approx 1000$.
\end{abstract}


\section{Introduction}

Recently there has been an increased interest in electronic structure
methods capable of accurately describing quasi-particle spectra and in
particular the ionization potential (IP) and electron affinity
(EA). Density function theory (DFT) has been the most commonly used
tool for predicting ground state properties for molecular and extended
systems.\cite{dreizler2012density,capelle2006bird,koch2015chemist,gidopoulos2003fundamentals}
Besides these properties, Kohn-Sham
(KS)~\cite{PhysRev.136.B864,kohn1965self} DFT offers a framework for
calculating the IPs from the orbital energy of the highest occupied
molecular orbital (HOMO), provided that exact exchange-correlation
functionals are
given.\cite{Perdew1982DFTdiscontinuities,cohen2008fractional} However,
in practice, the exact exchange-correlation functionals are not known,
and the IPs from KS-DFT are often off by several eVs in comparison to
experiments.\cite{seidl1996generalized}

The accurate description of quasi-particles has greatly benefited from
Green's functions techniques, mainly within the many-body perturbation
theory (MBPT).  These methods have proven extremely fruitful and allow
the inclusion of electron correlation through systematic
approximations of the self-energy, enabling an accurate description of
quasi-particle energies and lifetimes.  The most common flavor of
Green's function methods used is the GW
approximation,\cite{PhysRev.139.A796} where `G' indicates the
single-particle Green function and `W' the screened Coulomb
interaction.  This method offers improved accuracy over DFT in
describing quasiparticle properties including IPs and EAs in bulk
system.\cite{Hybertsen1985,Hybertsen1986,Rieger1999,Rinke2005,
  Neaton2006,Tiago2006,Friedrich2006,Gruning2006,Shishkin2007,
  Rostgaard2010,Tamblyn2011,Liao2011,Refaely-Abramson2011,Marom2012,Isseroff2012,Refaely-Abramson2012,Kronik2012}
In the GW approximation the contribution of exact exchange, while very
large, is only applied statistically.  The only dynamic part in the
approximation is based on RPA without exchange.  The validity of this
limitation when applied to molecular systems remains an active area of
research.\cite{Setten2015,vlcek2017stochastic}

An alternative to the GW approximation is the second-order self-energy
approximation, or the Green's function 2 (GF2) method, where the
self-energy is expanded to second-order in the Coulomb
interaction.\cite{cederbaum1975one,Holleboom-1990,stefanucci2013nonequilibrium,dahlen2005variational,phillips2014communication,pavovsevic2017communication,ohnishi2016explicitly,welden2015ionization,dahlen2005self,neuhauser2017stochastic}
In contrast to GW, GF2 includes exchange effects explicitly, beyond
the static-level, in the self-energy but treats the polarization term
differently than GW.  A key limitation of the GF2 method is the
$O(N_e^5)$ scaling of the second-order exchange term in the self
energy, rescricting its applications to small molecular systems.
Inspired by recent developments,\cite{neuhauser2017stochastic} we have
introduced a stochastic resolution of identity
(sRI)~\cite{Takeshita-2017stochastic} implementation of the Matsubara
GF2 approach for the calculation of ground state properties within the
second-order Green's function approach.\cite{takeshita2019stochastic}
The sRI technique reduced the computational cost of the second-order
self-energy method to $O(N_e^3)$, and was applied to systems with more
than a $1000$ electrons.

In the current work, we expand our approach and develop a stochastic
version of {\em real-time} GF2 theory. This provides a framework to
calculate quasi-particle properties, electron affinities, and
ionization potentials with a reduced scaling of $O(N_e^3)$.  To be
clear, in this work, we do not consider a time-dependent perturbation
potential and only propagate the single-particle Green's functions
along one real-time axis, i.e. we are considering only an equilibrium
scenario.  The stochastic real-time GF2 approach developed here has
similar flavor with previous stochastic version of electronic
structure theories,
e.g. MP2,\cite{neuhauser2012expeditious,ge2013guided,Takeshita-2017stochastic}
RPA,\cite{Rabani-2013-2}
DFT,\cite{Rabani-2013-3,Rabani-2014-4,cytter2014metropolis,chen2019overlapped}
and GW.\cite{PhysRevLett.113.076402}  Among the methods listed, this
work is closest to the stochastic implementation of Matsubara GF2
theory.\cite{takeshita2019stochastic} We illustrate the accuracy of
the approach for a set of molecules and compare the IPs to experiments
and GW results. We find that the stochastic real-time GF2 method
provides accurate IPs that are good agreement with experiments and
with the self-consistent GW method.

The manuscript is organized as follows: In Sec.~\ref{sec:theory} we
provide the basic theory for obtaining quasi-particle spectrum from
real time propagation of the second-order Green's function.  In
Sec.~\ref{sec:sRI} we review our stochastic resolution of identity and
apply such techniques to real-time GF2 theory. In
Sec.~\ref{sec:results} we report ionization potentials, quasi-particle
spectrum and timing from real-time GF2 theory. Finally, In
Sec.~\ref{sec:con}, we conclude.

\section{Theory} \label{sec:theory}
Consider a general Hamiltonian for a many-body electronic system in
second quantization
\begin{eqnarray} \label{eq:Hami}
\hat H = \sum_{ij} h_{ij}   \hat a^\dagger_i \hat a_j + \sum_{ijkl}v_{ijkl}  \hat a^\dagger_i \hat a^\dagger_k  \hat a_l \hat a_j,
\end{eqnarray} 
where $\hat a^\dagger_i$ ($\hat a_i$) is the creation (annihilation)
operator for an electron in atomic orbital $|\chi_i\rangle$. The
creation and annihilation operators obey the following commutation
relation:
\begin{eqnarray}
\left[\hat a_i ,\hat a^\dagger_j \right]=(S^{-1})_{ij}
\end{eqnarray}
Here $\mathbf{S}$ is the overlap matrix for
different orbitals, namely, $S_{ij} = \langle \chi_i | \chi_j
\rangle$.
In Eq.~(\ref{eq:Hami}), $h_{ij}$ are matrix elements of the non-interacting
electronic Hamiltonian and $v_{ijkl}$ are the 4-index electron
repulsion integrals:
\begin{equation} 
	\label{eq:2e4c}
	v_{ijkl} = (i  j | kl ) = \iint d{\bf r}_1 d{\bf r}_2 \frac{\chi_i ({\bf r}_1)\chi_j ({\bf r}_1)\chi_k ({\bf r}_2)\chi_l ({\bf r}_2)}{\left|{\bf r}_1-{\bf r}_2\right|},
\end{equation}
where $\chi_i ({\bf r})$ is the position representation of
$|\chi_i\rangle$.

\subsection{Kadanoff-Baym Equations}
The quantity of interest in this work is the single-particle Green's
function on the Keldysh contour, defined as (we set $\hbar=1$
  throughout):\cite{haug2008quantum,stefanucci2013nonequilibrium}
\begin{eqnarray}
G_{ij} (\tilde{t}_1, \tilde{t}_2)  = -i \langle T_C \hat a_i
(\tilde{t}_1) \hat a^\dagger_j (\tilde{t}_2) \rangle.
\label{eq:Gij}
\end{eqnarray}
We use $\tilde{t}$ to denote a time point on a Keldysh contour defined
on the real axis from $0$ to positive infinity $(0,+\infty)$, then
back to origin $(+\infty, 0)$, finally to $-i\beta$ on imaginary axis
$(0, -i\beta)$. $T_C$ is a time ordering operator on the Keldysh
contour.  The operators in the above equations are defined in the
Heisenberg representation such that $\hat a^\dagger_i (\tilde{t}_1) =
e^{i \hat H \tilde{t}_1} \hat a^\dagger_i e^{-i \hat H \tilde{t}_1}$.
The average in the above is taken with respect to a Boltzmann
distribution: $\langle \cdots \rangle = Z^{-1} \mbox{Tr}
\left[(\cdots) e^{-\beta (\hat H - \mu \hat N) } \right]$, where $Z =
\mbox{Tr}\left[e^{-\beta (\hat H - \mu \hat N)}\right]$ is the
grand-canonical partition function, $\beta=\frac{1}{k_{\rm B}T}$
is the inverse temperature, and $\mu$ is the chemical potential. The
number operator is given by $\hat N = \sum_{ij} S_{ij} \hat
a^\dagger_i \hat a_j $.

\begin{figure}[htbp] 
   \centering
   \includegraphics[width=3in]{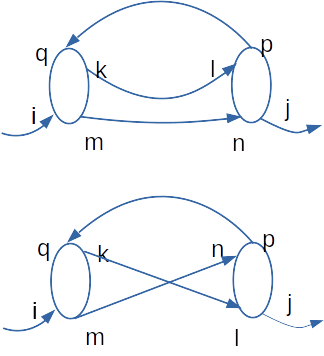} 
   \caption{second-order Born self energy for molecules: direct
     (upper) and exchange (lower) correlations. See also
     Eq.~(\ref{eq:sigma-2}). Note that the exchange correlations (lower)
     are not included in GW approximation.}
   \label{fig:self-energy}
\end{figure}

The equation of motion for the Green's function defined in
Eq.~(\ref{eq:Gij}) satisfies the Kadanoff-Baym equation:
\begin{eqnarray}
i\mathbf{S} \partial_{\tilde{t}_1} \mathbf{G}(\tilde{t}_1, \tilde{t}_2) = \delta(\tilde{t}_1, \tilde{t}_2) + \mathbf{F}\mathbf{G}(\tilde{t}_1, \tilde{t}_2) + \int_C \mathbf{\Sigma} (\tilde{t}_1, \tilde{t}_3) \mathbf{G} (\tilde{t}_3, \tilde{t}_2 )d\tilde{t}_3,
\end{eqnarray}
where $\mathbf{F}$ is the Fock matrix obtained from the imaginary-time
Matsubara Green's function (see below for more details) and the time
integration in the above equation is carried out on the Keldysh
contour ($C$). In the second-order Born approximation, the matrix
elements of the self-energy $\mathbf{\Sigma}(\tilde{t}_1,
\tilde{t}_2)$ take the following form:
\begin{equation}
\label{eq:sigma-2}
\Sigma_{ij} (\tilde{t}_1, \tilde{t}_2) =
\sum_{klmnpq}G_{kl}(\tilde{t}_1, \tilde{t}_2)G_{mn}(\tilde{t}_1,
\tilde{t}_2)G_{pq}(\tilde{t}_2, \tilde{t}_1)v_{imqk}(2v_{lpnj} -
v_{nplj}).
\end{equation}
In contrast to the GW approximation, in the second-order Born
approximation, the exchange correlations are taken into accounted
explicitly beyond the static part (see Fig. \ref{fig:self-energy}). The factor $2$ in the
above equation accounts for spin degeneracy. Note that for simplicity
we have restricted ourselves to the closed-shell case, but extensions
to open-shell systems are straightforward.

To solve the Kadanoff-Baym equations requires a specific projection
onto real and imaginary time branches of the Keldysh contour. In the
present case (equilibrium), this requires only three types of GFs:
When both times are projected onto the imaginary branch (Matsubara
GF), when one time is projected onto the imaginary-time branch while
the other is projected onto the real-time branch (mixed-time GF), and
finally, when both times are projected to the real-time branch for
$t,t'>0$ (retarded GF) in order to obtain the spectral function.  In
the following subsections we describe equations of motion for the
three cases discussed above.  We begin with the simplest case where
both times are projected onto the imaginary axis.

\subsection{Matsubara Green's function} \label{subsec:MGF}
If we restrict $\tilde{t}_1$
and $\tilde{t}_2$ to the imaginary time branch ($\tilde{t}_1 =
-i\tau_1, \tilde{t}_2 = -i\tau_2$), the Keldysh contour-ordered
Green's function in Eq.~(\ref{eq:Gij}) reduces to the Matsubara
Green's function, $i \mathbf{G}^M(\tau_1 - \tau_2) =
\mathbf{G}(-i\tau_1, -i\tau_2)$, which only depends on the
imaginary-time difference ($\tau = \tau_1 - \tau_2 \in
[0,\beta]$). The superscript ``M'' stands for Matsubara quantity.  The
equation of motion for the Matsubara Green's function can be written
in an integral form:
\begin{eqnarray} \label{eq:MatG}
\mathbf{G}^M (\tau)=
          \mathbf{G}^M_0(\tau) + \int_0^\beta d\tau' d\tau''
          \mathbf{G}^M_0(\tau-\tau') \mathbf {\Sigma}^M (\tau'-\tau'')
          \mathbf{G}^M (\tau''),
\end{eqnarray}
where $\mathbf{G}^M_0(\tau)$ is the zeroth order Matsubara Green's
function given in term of the Fockian:~\cite{takeshita2019stochastic}
\begin{eqnarray} 
 \mathbf{G}^M_0(\tau) = \mathbf X
        e^{- \tau (\mathbf {\bar F} -\mu \mathbf {I} )} \left[ \frac{\theta(-
            \tau)}{1+ e^{\beta(\mathbf {\bar F} - \mu \mathbf {I} )}} -
          \frac{\theta( \tau)}{1+ e^{- \beta(\mathbf{\bar F} - \mu
              \mathbf{I} )} } \right] \mathbf X^T.
\end{eqnarray}
In the above equation, $\mathbf{X} \mathbf{X}^T = \mathbf{S}^{-1} $
and $\mathbf{\bar F} =\mathbf{X}^{T} \mathbf{F} \mathbf{X}$.  In the
second-order Born approximation, the matrix elements of the Matsubara
self-energy take the following form:
\begin{equation}
\label{eq:MatSig}
	\Sigma^M_{ij} (\tau) =
\sum_{klmnpq}G^M_{kl}(\tau)G^M_{mn}(\tau)G^M_{pq}(\beta-\tau)v_{imqk}(2v_{lpnj} - v_{nplj}).
\end{equation}

Obviously, Eqs.~(\ref{eq:MatG})-(\ref{eq:MatSig}) have to be solved
self-consistently, since the self-energy itself depends on the
Matsubara GF.  Furthermore, for convenience we also update the Fock
matrix according to~\cite{takeshita2019stochastic} $F_{ij} = h_{ij} -2
\sum_{mn} G_{mn} ( \tau = \beta) (v_{ijmn} - \frac12 v_{inmj})$ and
also adjust the chemical potential $\mu$ to conserve the number of
electrons by imposing that $N_e = -2 \sum_{mn} G_{mn} ( \tau = \beta)
S_{mn} $.  In order to solve for the Matsubara Green's function in a
numerically efficient way, proper quadratures and contractions are
used to evaluate the double integral on the right hand side of
Eq.~(\ref{eq:MatG}). See Ref.~\citen{takeshita2019stochastic} for more
details.

\subsection{Equations of motion for mixed-time Green's function}
In the absence of a time-dependent perturbation, it is sufficient to
work with the mixed branch, such that the mixed-time Green's function
is given by $\mathbf{G}^\rceil(t, \tau ) = \mathbf{G} (\tilde{t}_1 =
t, \tilde{t}_2 = -i \tau)$. Using Langreth
rules,\cite{stefanucci2013nonequilibrium} the equation of motion for
the mixed-time Green's function can be written as
\begin{eqnarray} \label{eq:Grceil}
i \mathbf{S} \partial_{t} \mathbf{G}^\rceil(t, \tau ) =  \mathbf{F} \mathbf{G}^\rceil (t, \tau ) + \int_{0}^{t} \mathbf{\Sigma}^R(t') \mathbf{G}^\rceil (t - t', \tau )dt' 
+ \int_0^\beta \mathbf{\Sigma}^\rceil ( t, \tau_1) \mathbf{G}^M ( \tau_1- \tau) d\tau_1.
\end{eqnarray}
In the above equation, both ${\mathbf F}$ and $\mathbf{G}^M ( \tau_1-
\tau)$ are obtained from the solution of the Kadanoff-Baym equations
in imaginary time as explained in the previous subsection. In other
words, the matrix elements of ${\mathbf F}$ are given by $F_{ij} =
h_{ij} -2 \sum_{mn} G_{mn} ( \tau = \beta) (v_{ijmn} - \frac12
v_{inmj})$. The mixed-time self-energy within the second Born
approximation is given by:
\begin{eqnarray} \label{eq:Sigceil}
\Sigma^\rceil_{ij}(t, \tau ) = \sum_{klmnpq} G^\rceil_{kl}(t, \tau ) G^\rceil_{mn}(t, \tau ) G^\rceil_{pq}(t, \beta-\tau )^* v_{impk} (2 v_{jnql} - v_{jlqn}),
\end{eqnarray}
where we have used the relation $G^\lceil_{ji}(\tau, t )^*=
G^\rceil_{ij}( t, \beta -\tau)$ and defined $\mathbf{G}^\lceil(\tau, t
) = \mathbf{G} (\tilde{t}_1 = -i\tau, \tilde{t}_2 = t)$ to be
consistent with the definition of $\mathbf{G}^\rceil(t, \tau )$.
Finally, the retarded self-energy $\Sigma^R_{ij}(t_1)$ is related to
lesser and greater self-energies by the simple relation:
\begin{eqnarray}
  \Sigma^R_{ij}(t_1-t_2) = \theta(t_1-t_2) (\Sigma^>_{ij}(t_1-t_2) - \Sigma^<_{ij}(t_1-t_2) ).
  \label{eq:SigR}
\end{eqnarray}
where $\theta(t)$ is the Heaviside step function and
$\Sigma^{<,>}_{ij}(t)$ are the matrix elements of the lesser/greater
self-energy. At equilibrium, the latter can be expressed in terms of
the mixed-time self-energies:
\begin{eqnarray} \label{eq:Sigma<}
\Sigma^<_{ij}(t) &=& \Sigma^\rceil_{ij}(t, \tau = 0 ) \\
\Sigma^>_{ij}(t) &=& - \Sigma^>_{ji}(-t)^* = - \Sigma^\lceil_{ji}(\tau=0, t )^*=- \Sigma^\rceil_{ij}( t, \tau = \beta ) \label{eq:Sigma>}
\end{eqnarray}

The equation of motion for $\mathbf{G}^\rceil(t, \tau )$ must be
solved self-consistently, since the corresponding self-energies
depends on the GF itself.  The initial conditions for the mixed-time
GF in Eq.~(\ref{eq:Grceil}) is given by $G_{ij}^\rceil (t=0, \tau) = -
i G_{ij}^M (\beta - \tau)$.  In the numerical implementations, we have
used the method of Ref.~\citen{stan2009time} to propagate the
mixed-time GFs. Similar to the case of pure imaginary time, the last
term in Eq.~(\ref{eq:Grceil}) is evaluated through proper
quadrature.\cite{takeshita2019stochastic}

\subsection{Observables and Quasi-particle Spectrum}
In order to calculate the spectral functions within the second-order
Green's function approach, one requires the lesser and greater Green's
functions.  These can be obtained directly from $G^\rceil(t_1, \tau )$
as follows (relation holds for equilibrium only):
\begin{eqnarray} \label{eq:G<}
G^<_{ij}(t) &=& G^\rceil_{ij}(t, \tau = 0 ) \\
G^>_{ij}(t) &=& - G^>_{ji}(-t)^* = -G^\lceil_{ji}(\tau=0, t )^*=-G^\rceil_{ij}( t, \tau = \beta ).
\end{eqnarray}
Furthermore, the retarded Green's function can be expressed in terms
of the lesser/greater GFs as follows:
\begin{eqnarray}
G^R_{ij}(t) = \theta(t) (G^>_{ij}(t) - G^<_{ij}(t) ),
\end{eqnarray}
and spectral function $A(\omega)$ is then defined as the imaginary
portion of the retarded Green's function
\begin{eqnarray} \label{eq:Aomega}
A(\omega) =  - \sum_{mn} \mbox{Im} \tilde{G}^R_{mn}(\omega) {S}_{mn}
\end{eqnarray}
where $\tilde{\mathbf{G}}^R(\omega)$ is the Fourier transform of $\mathbf{G}^R(t)$, 
\begin{eqnarray} \label{eq:fftGr}
\tilde{\mathbf{G}}^R(\omega) = \int_{-\infty}^\infty dt \mathbf{G}^R(t)e^{i\omega t} 
\end{eqnarray}
As can be clearly seen, the spectral function can be obtained directly
from the mixed-time $G^\rceil_{ij}( t, \tau )$.

\section{Stochastic Resolution of Identity} \label{sec:sRI}
Similar to the Matsubara GF2 case, the computational bottleneck in
real-time propagation of the Green's function is the evaluation of the
self-energy in Eq.~(\ref{eq:Sigceil}), which scales as $O(N_e^5)$. To
overcome this steep computational scaling, we have developed a
stochastic resolution of identity (sRI) for Matsubara GF2 theory,
which reduces the computational cost of the self-energy to
$O(N_e^3)$. \cite{Takeshita-2017stochastic,takeshita2019stochastic}
The same technique is applied here to the mixed-time formulation. In
this section, we briefly review the sRI theory and show how sRI
formulation can be used to reduce the computational cost in evaluation
of the mixed-time self-energy.

Before introducing sRI, we first review the RI. The 4-index electron
repulsion integral (ERI) defined in Eq.~(\ref{eq:2e4c}) can be
approximated by
	\begin{equation}
	 \label{eq:abgd-stoch}
	\begin{split}
		(i  j | m n)  &\approx \sum_{AB}^{N_{\rm aux}} (i j |A)  V^{-1}_{AB} (B| mn ) 
	\end{split}
	\end{equation}
where we have defined the 3-index ERI and 2-index ERI as following:
\begin{equation}
		(i j |A) = \iint dr_1 dr_2 \frac{\chi_i (r_1)\chi_j (r_1)\chi_A(r_2)}{r_{12}}
	\end{equation}
	\begin{equation}
		V_{AB} = \iint dr_1 dr_2 \frac{\chi_A(r_1) \chi_B(r_2)}{r_{12}}.
	\end{equation}
Here, $\chi_A$ and $\chi_B$ are auxiliary orbitals. 

In a stochastic resolution of identity approach, an additional set of
$N_s$ \textit{stochastic orbitals} are introduced, $\{ \theta^\xi \}$,
$\xi = 1, 2, \cdots, N_s$. Here $\theta^\xi$ is a vector of length
$N_{\rm aux}$ ($N_{\rm aux}$ is the size of the auxiliary basis). The
elements in $\theta^\xi$ are randomly chosen from a uniform
distribution of $\pm1$, $\theta^\xi_A = \pm1$ ($A=1,2,\cdots N_{\rm
  aux}$) and satisfy the relation
	\begin{equation}
		\begin{split}
	 	  \lim_{N_s \rightarrow \infty}\frac{1}{N_s} \sum_{\xi}^{N_s} \theta^\xi_A \theta^\xi_B = \delta_{AB} ,
		\end{split}
	\end{equation}
Using the stochastic orbitals, Eq.~(\ref{eq:abgd-stoch}) can be expressed as follows:
\begin{eqnarray}
	 \label{eq:abgd-stoch1}
		\sum_{AB}^{N_{\rm aux}} (i j |A)  V^{-1}_{AB} (B| mn )  &=&  \sum_{PQ}^{N_{aux}} \sum_{AB}^{N_{\rm aux}} (i j |A)V^{-\frac{1}{2}}_{AP} \delta_{PQ}  V^{-\frac{1}{2}}_{QB} (B| mn ) \nonumber \\
		&\rightarrow&  \frac{1}{N_s} \sum_{\xi}^{N_s} \sum_{PQ}^{N_{\rm aux}} \sum_{AB}^{N_{\rm aux}} ( ij  |A)V^{-\frac{1}{2}}_{AP} \theta^\xi_P \theta^\xi_Q V^{-\frac{1}{2}}_{QB} (B| mn ) \nonumber \\
		& = & \frac{1}{N_s} \sum_{\xi}^{N_s}  \left[ \sum_{A}^{N_{\rm aux}} ( ij |A) \sum_P^{N_{\rm aux}}V^{-\frac{1}{2}}_{AP}  \theta^{\xi}_{P} \right] \left[  \sum_{B}^{N_{\rm aux}}  (B| mn )  \sum_Q^{N_{\rm aux}} \theta_{Q}^{\xi}   V^{-\frac{1}{2}}_{QB} \right] 
\end{eqnarray}
The 4-index ERI (Eq.~\ref{eq:abgd-stoch}) now can be approximated by
an average over number of stochastic orbitals,
	\begin{equation}
	\label{eq:sRI}
	( i j  | mn) \approx \frac{1}{N_s} \sum_\xi  R_{i j }^{\xi} R_{mn}^{\xi} \equiv \left< R_{ij} R_{mn}  \right>_\xi,
	\end{equation}
where $R_{ i j }^{\xi}$ is given by:
\begin{equation}
	\label{eq:4}
         R_{ i j }^{\xi} = \sum_A^{N_{\rm aux}} ( i j |A)\left[ \sum_P^{N_{\rm aux}} [ V^{-\frac{1}{2}}_{AP} \theta_P^{\xi} ] \right].
\end{equation}

The advantages of introduction of sRI over RI have been discussed
extensively in our previous
work.\cite{Takeshita-2017stochastic,takeshita2019stochastic}
Particularly, sRI reduces the overall computational scaling since the
number of stochastic orbital does not increases with system size.  The
current focus of this paper is to apply the sRI method to our
real-time GF2 theory.

\subsection{sRI applied to real-time second Born approximation}
In the current formulation, we apply the sRI to the second Born
approximation for the real-time self-energy, which takes the following form:
\begin{equation}
\begin{split}
	\Sigma^\rceil_{ij}(t_1, \tau ) =&\left< \sum_{klmnpq} G^\rceil_{kl}(t_1, \tau ) G^\rceil_{mn}(t_1, \tau ) G^\rceil_{pq}(t_1, \beta-\tau )^* R_{ik}R_{mq} (2
        R^{\prime}_{lj}R^{\prime}_{pn} -
        R^{\prime}_{nj}R^{\prime}_{lp})\right>_{\zeta,
          \zeta^{\prime}}\\ =&\Big< \sum_{klmnpq} 2 G^\rceil_{kl}(t_1, \tau ) G^\rceil_{mn}(t_1, \tau ) G^\rceil_{pq}(t_1, \beta-\tau )^* R_{ik}R_{mq}
        R^{\prime}_{lj}R^{\prime}_{pn} \\ & - G^\rceil_{kl}(t_1, \tau ) G^\rceil_{mn}(t_1, \tau ) G^\rceil_{pq}(t_1, \beta-\tau )^*R_{ik}R_{mq} R^{\prime}_{nj}R^{\prime}_{lp}
        \Big>_{\zeta, \zeta^{\prime}} \\ =& 
        \Sigma_{ij}^{\rm dir,\rceil} (t, \tau) + \Sigma_{ij}^{\rm ex,\rceil} (t, \tau) 
\end{split} \label{eq:sRI-SE}
\end{equation}
Here $R$ and $R'$ are uncorrelated stochastic matrices generated using
a different set of stochastic orbitals (Eq.~(\ref{eq:sRI})) and the
direct and exchange terms of the self-energies are given by:
\begin{equation}
\begin{split}
	 \Sigma_{ij}^{\rm dir,\rceil} (t, \tau) =&\Big< \sum_{klmnpq} 2 G^\rceil_{kl}(t_1, \tau ) G^\rceil_{mn}(t_1, \tau ) G^\rceil_{pq}(t_1, \beta-\tau )^*
         R^{\prime}_{lj}R^{\prime}_{pn} \Big>_{\zeta, \zeta^{\prime}}
         \\ \Sigma_{ij}^{\rm ex,\rceil} (t, \tau)  =&\Big< -\sum_{klmnpq}  G^\rceil_{kl}(t_1, \tau ) G^\rceil_{mn}(t_1, \tau ) G^\rceil_{pq}(t_1, \beta-\tau )^* R_{ik}R_{mq}
         R^{\prime}_{nj}R^{\prime}_{lp} \Big>_{\zeta,
           \zeta^{\prime}}\\
\end{split} \label{eq:sRI-SE2}
\end{equation}
The above expressions for the mixed-time self-energy can be evaluated
at $O(N_e^3)$ computational scaling (rather than $O(N_e^5)$) as long
as the number of stochastic orbitals does not increase with the system
size.  This can be done by using contractions. Note that the sRI is
used only for the evaluation of the self-energy while the remaining
portion of the calculations is performed deterministically.

\subsection{Summary of the proposed algorithm}
To summarize this part, the real-time sRI-GF2 requires the following
steps:
\begin{enumerate}
\item Perform a sRI-GF2 for the Matsubara Green's function as
  described in subsection \ref{subsec:MGF} and in more detail in
  Ref.~\citen{takeshita2019stochastic} to generate ${\bf G}^M(\tau)$.

\item Use ${\bf G}^M(\tau)$ as the initial condition for the
  mixed-time Green's function and solve Eq.~(\ref{eq:Grceil} and
  obtain the mixed-time self-energy using Eq.~(\ref{eq:Sigceil}).

\item Solve Eq.~(\ref{eq:Grceil}) to generate the mixed-time
  self-energy and Eqs.~(\ref{eq:SigR})-(\ref{eq:Sigma>}) to generate
  the retarded self-energy, $\Sigma^R_{ij}(t)$.  This is done ``on the fly''.

\item Propagate Eq.~(\ref{eq:Grceil})until the self-energies decay to
  a predefined tolerance or until the final observable converge with
  respect to the propagation time.

\item Fourier transform $\Sigma^R_{ij}(t)$ to the frequency domain and
  solve the Dyson equation for $\tilde{\mathbf{G}}^R(\omega) =
  \frac{1}{\omega \mathbf{S} - \mathbf{F} -
    \tilde{\mathbf{\Sigma}}^R(\omega)}$.  Use Eq.~(\ref{eq:Aomega}) to
  generate the spectral function, $A(\omega)$. Here,
  $\tilde{\mathbf{\Sigma}}^R(\omega)$ is the Fourier transform of
    $\mathbf{\Sigma}^R(t)$.
\end{enumerate}

\section{Results and Discussion} \label{sec:results}
In this section, we analyze the performance of the real-time sRI-GF2
theory, especially its ability to predicting IPs and quasi-particle
spectra for molecules and for extended systems. The time step used to
integrate Eq.~(\ref{eq:Grceil}) is $0.05 E_h^{-1}$ ($E_h$ is Hartree
energy),\cite{vlcek2017stochastic} unless otherwise noted. We set
$\beta = 50 E_h^{-1}$ and use $256$ Chebyshev points and
Gauss-Legendre quadratures to integrate the imaginary term in
Eq.~(\ref{eq:Grceil}). In addition, a small damping term $\eta=0.01
E_h$ is added to the real-time propagation of the Green's function.
Finally, a complementary error function erfc($t/t_{ce}$) is multiplied
to $\Sigma^R (t)$ in order to prevent instability of Fourier
transform.\cite{neuhauser2012expeditious,noteForFFT}.

\subsection{Ionization potentials for molecules}
The IP can be extracted from the quasi-particle
spectrum $A(\omega)$ as the position of the peak near the highest
occupied molecular orbital (HOMO). In this subsection, we compare IPs
generated from the real-time sRI-GF2 to IPs from Hartree-Fock (HF),
$G_0W_0$, and fully self-consistent $GW$ (SCGW) for a set of
molecules.  In HF theory, the IP is given by the HOMO energy as
suggested by Koopmans' theorem. The results for $G_0W_0$ and SCGW
performed over HF are taken from Ref.~\citen{PhysRevB.89.155417}.  The
basis set chosen here is cc-pvdz.  We have used basis def2-qzvp-ri for
fitting ERI (Eq.~(\ref{eq:abgd-stoch})) and def2-qzvp-jkfit for
fitting Fock matrix.

\begin{figure}[htbp] 
   \centering
   \includegraphics[width=4in]{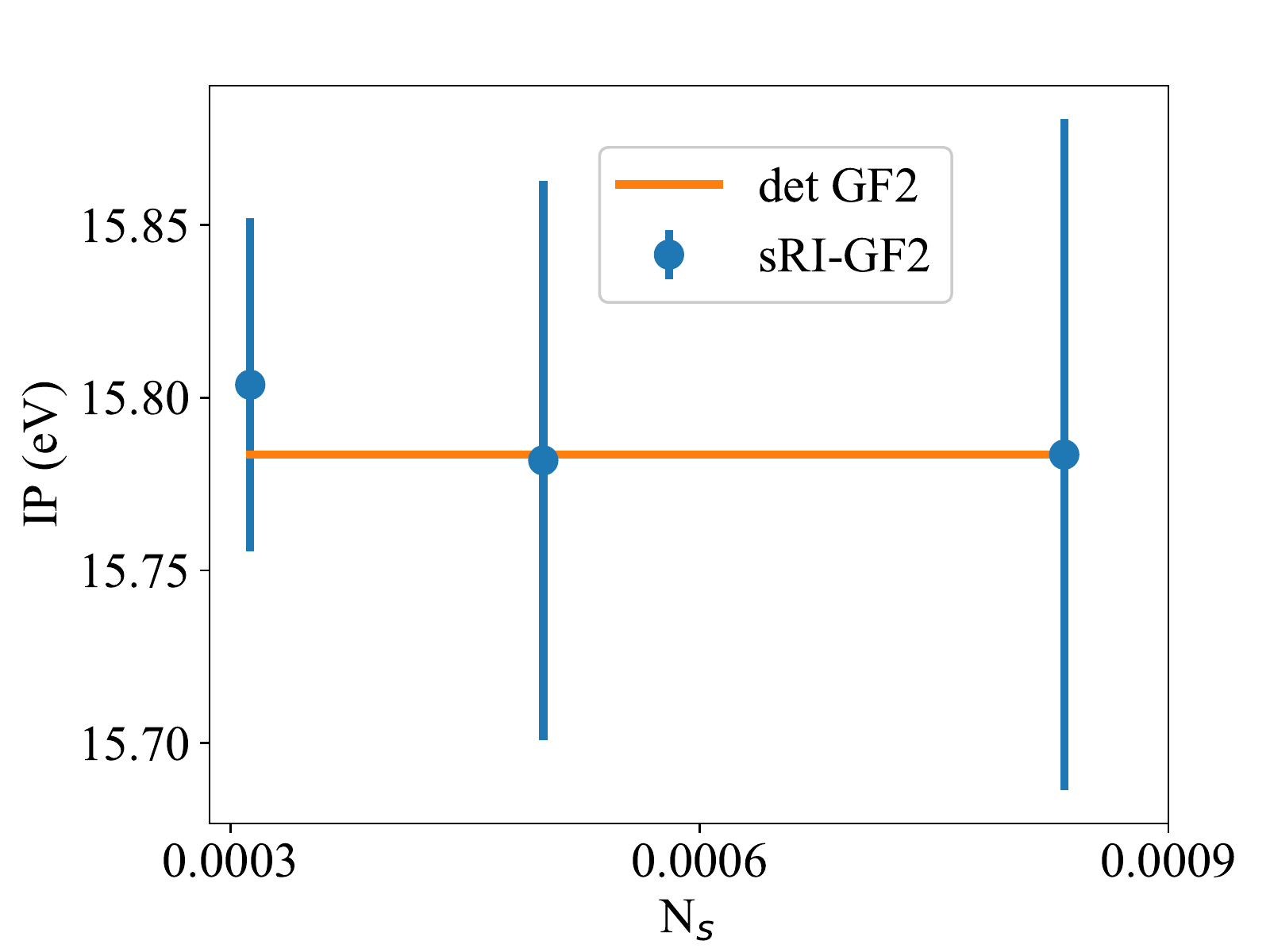} 
   \caption{IP for H$_2$ molecule from sRI-GF2 and deterministic
     GF2. In sRI-GF2 calculations, we use $N_s$=1200,
     2000, 3200 stochastic orbitals. The stochastic error decreases
     when increasing number of stochastic orbitals. }
   \label{fig:error}
\end{figure}

We first benchmark our real-time sRI-GF2 results against deterministic
GF2 results. In Fig.~\ref{fig:error}, we plot IP for H$_2$ molecule
from sRI-GF2 results using $N_s$=1200, 2000, 3200
stochastic orbitals. The errors in sRI-GF2 results are estimated by
the standard deviation of 10 independent runs using different
seeds. As expected, when increasing the number of stochastic orbitals,
the errors in sRI-GF2 results decreases. Note that sRI-GF2 predicts
IPs in excellent agreement with deterministic GF2 within the error
bar. 
   
\begin{table} 
\caption{Ionization potentials (in eV) for a list of atoms and  molecules}
\begin{tabular}{ c c c c c c c c} 
 \hline \hline
             &   Exp      &HF      &  $G_0W_0$  &  SCGW  & sRI-GF2   & sRI-GF2 no DE & DE effect\\ \hline 
He         &  24.59     &24.88       &      24.36   & 24.28  & 24.01$\pm$ 0.02  &\\ 
Be         &  9.32     &8.41      &     8.98  & 8.46 & 8.23 $\pm$ 0.02 &\\
Ne         &  21.56     &22.64         &   20.87  & 20.98 & 21.44 $\pm$ 0.09 &\\ 
H$_2$   &  15.43     &16.11      &   16.23   & 16.00 & 15.78 $\pm$ 0.03 & 15.74$\pm$ 0.04 & 0.04\\ 
CH$_4$&   13.60     &14.78        &   14.43   & 14.09 & 13.24 $\pm$ 0.10 & 13.16$\pm$0.11 & 0.08\\ 
LiH        &   7.90    &8.18        &   7.96   & 7.74 & 7.73 $\pm$ 0.02 & 7.45 $\pm$ 0.05 & 0.28\\ 
LiF        &   11.3     &12.66      &   10.72   & 10.85 & 10.77 $\pm$ 0.07 & 9.83 $\pm$ 0.10 & 0.94\\ 
HF        &   16.12     &17.11        &   15.55  & 15.54 &15.19 $\pm$ 0.13 & 15.01$\pm$ 0.19 & 0.18\\ 
H$_2$O &  12.62      &13.42        &  12.17  & 12.03 & 11.92 $\pm$ 0.10 & 11.80 $\pm$ 0.09 & 0.12 \\ \hline
Error  &  0.00       & 0.84        &  0.506   & 0.51  & 0.537 &\\ \hline\hline
\end{tabular} \label{table:1}
\end{table}

In Table~\ref{table:1}, we list the IPs for a set of selected atoms
and molecules. We provide experimental results as well as calculated
IPs using HF, $G_0W_0$ and SCGW and compare these to the real-time
sRI-GF2 approach developed herein.  The results from sRI-GF2 are mean
values over of independent runs. The errors from sRI-GF2 are estimated
by the standard deviations of the mean values. We also provide the
mean absolute error against experimental results.  Overall, we find
very good agreement between the many-body perturbation techniques
based on the GW approximation and the real-time sRI-GF2 approach.  The
performance of GF2 is comparable and sometimes better than the
many-body perturbation technique within the GW approximation. This is
a significant observation, since GW is considered the state of the art
for describing IPs, even for small
molecules.\cite{Setten2015,vlcek2017stochastic} This suggests that,
for some molecules, a reliable estimate of the IPs can be obtained
within a theoretical framework of GF2 without the need to compute
screened Coulomb interactions. Note that in GF2, the dynamic exchange
term is included explicitly in self-energy (see
Fig.~\ref{fig:self-energy}).  To examine the role of dynamic exchange
on the IPs, in Table~\ref{table:1} we list IPs for a set of molecules
where the dynamic exchange (DE) term in the self-energy was absent. We
refer to this as ``sRI-GF2 no dynamic exchange (DE)''.  Note that
sRI-GF2 no DE tends to under-estimate IPs as compared to sRI-GF2 and
experiments. The last column in Table~\ref{table:1} shows the
difference in IPs between sRI-GF2 and sRI-GF2 no exchange.  These
results suggest that the contribution of the dynamic exchange term is
significant, up to $\approx 1$~eV corrections to the IPs.

\begin{figure}[htbp] 
   \centering
   \includegraphics[width=4in]{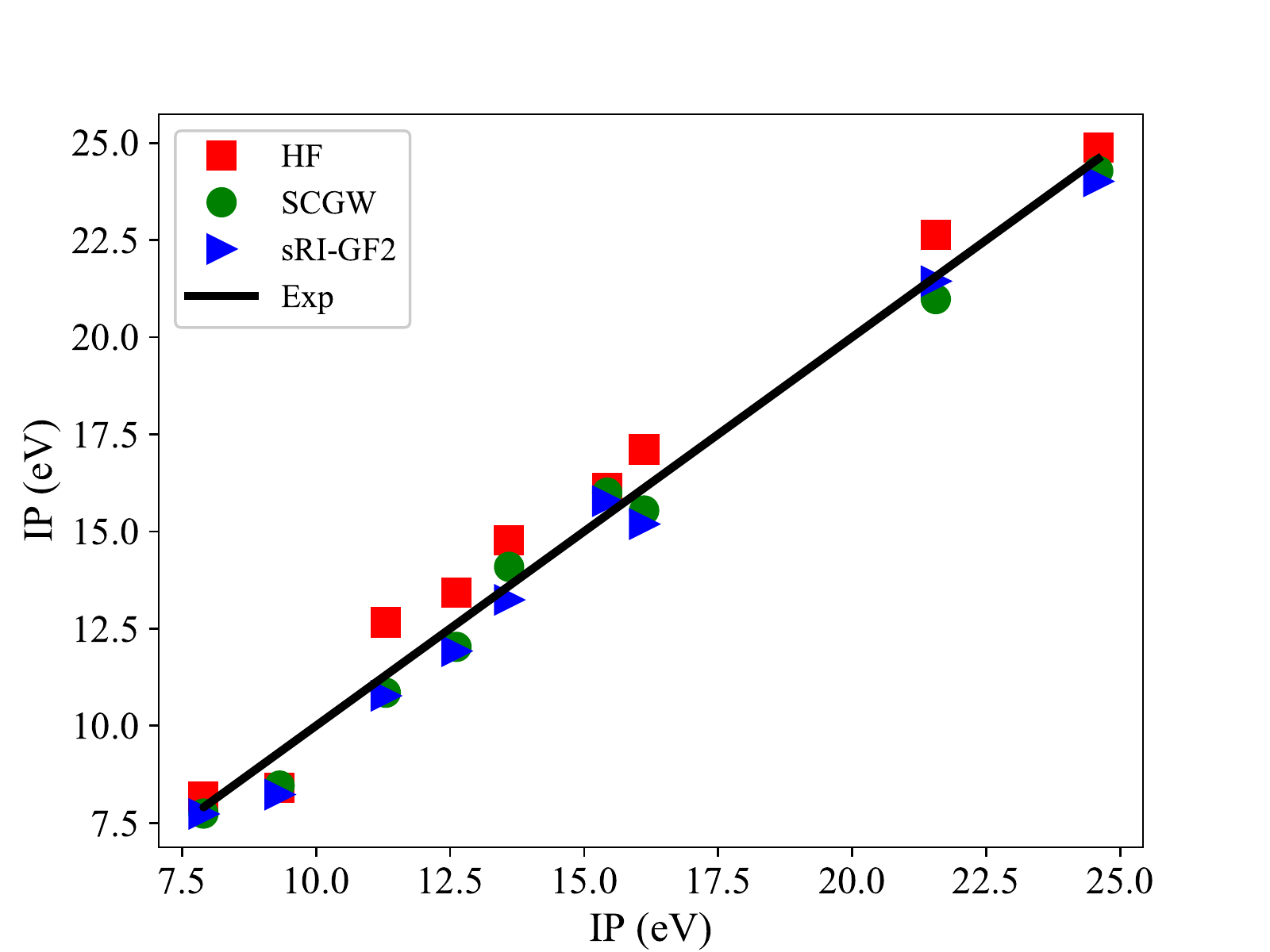} 
   \caption{Comparison of IPs calculated from HF, SCGW, GF2 against
     experimental results for molecules listed in Table~\ref{table:1}.
     Note that HF tends to over-estimate the IP. SCGW and GF2 have
     similar accuracy in predicting IP.}
   \label{fig:IP}
\end{figure}

Fig.~\ref{fig:IP} provides a more compelling illustration of the
results summarized in Table.~\ref{table:1}. The horizontal axis in
Fig.~\ref{fig:IP} is the experimental IP and the vertical axis is the
IP calculated by the different methods. We find that HF, as expected,
overestimates the IP for most molecules studied here. By incorporating
electron correlations, the GW and GF2 methods provide much better
agreement with experiments.

\subsection{Spectral functions and computational complexity: Hydrogen dimer chains}
While the IP for molecular systems can be obtained using an extended
Koopmans' theorem,\cite{welden2015ionization} such an theorem cannot
be used to explore the entire frequency range of the spectral
function. This brings us to one of the main advantages of the
time-domain formalism: With the same computational costs to obtain the
IPs, we can also calculated the quasiparticle spectrum over a wide
range of frequencies.  To demonstrate this within the real-time
sRI-GF2, we have carried calculations for the quasi-particle spectrum
of a series of hydrogen dimer chains of different length, $N$. We
demonstrate that the stochastic RI approach allows us to extent the
size of systems that can be described within GF2 with scaling that is
slightly better than $O(N_e^3)$.

\begin{figure}[htbp] 
   \centering
   \subfloat[H$_{20}$]{\includegraphics[width=3.5in]{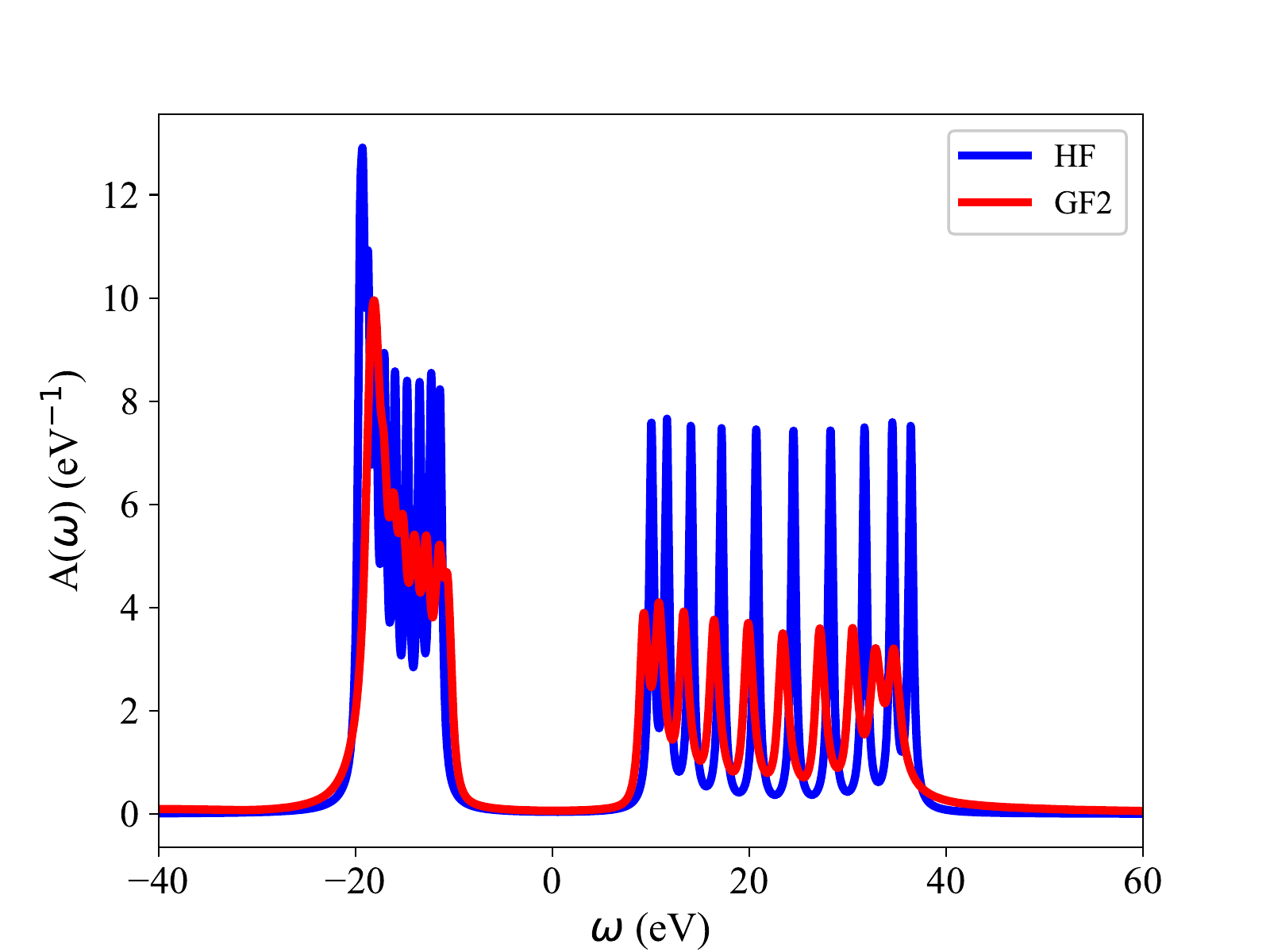} } 
   \subfloat[H$_{80}$]{\includegraphics[width=3.5in]{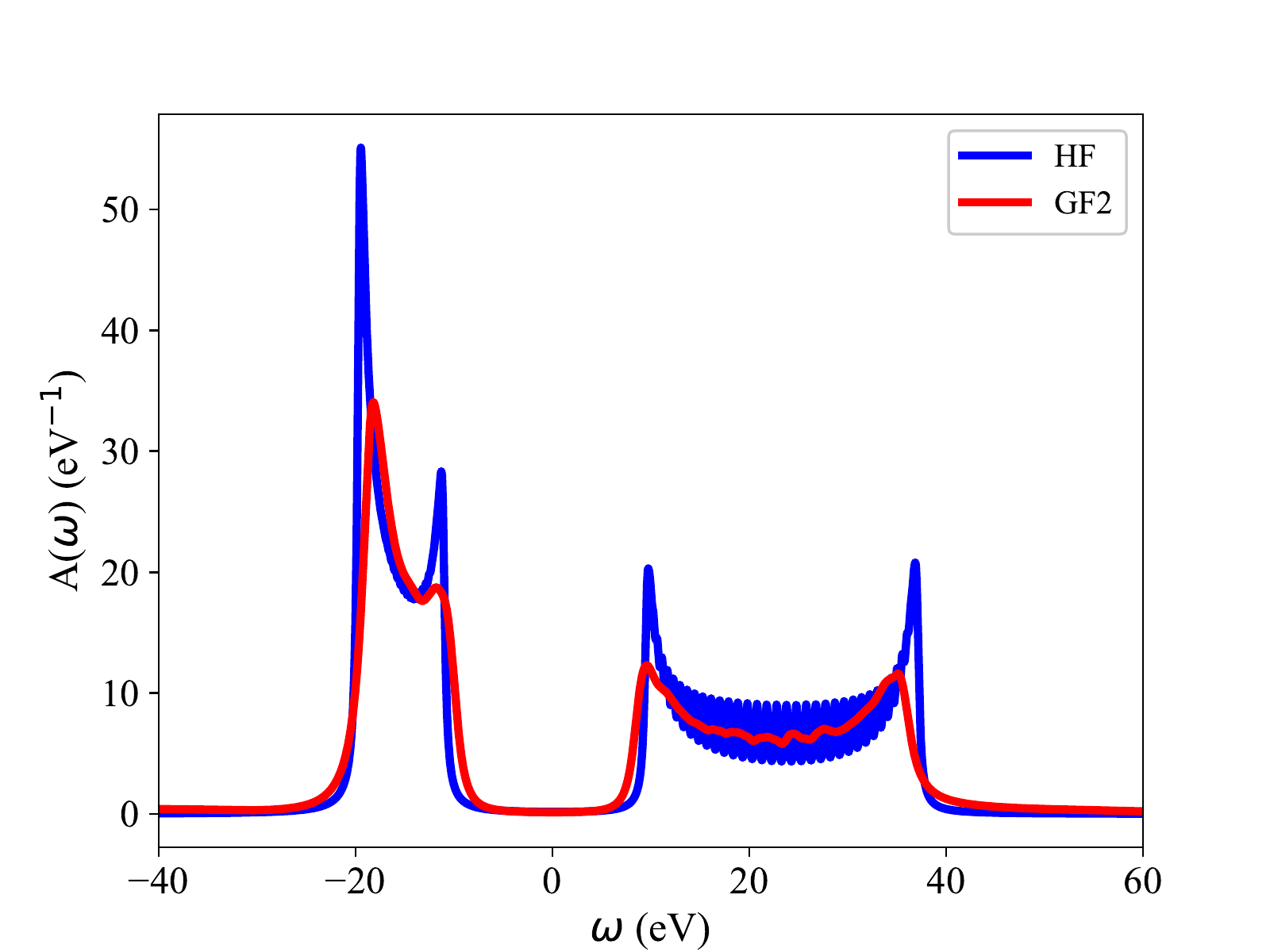} }
   
   \subfloat[H$_{200}$]{\includegraphics[width=3.5in]{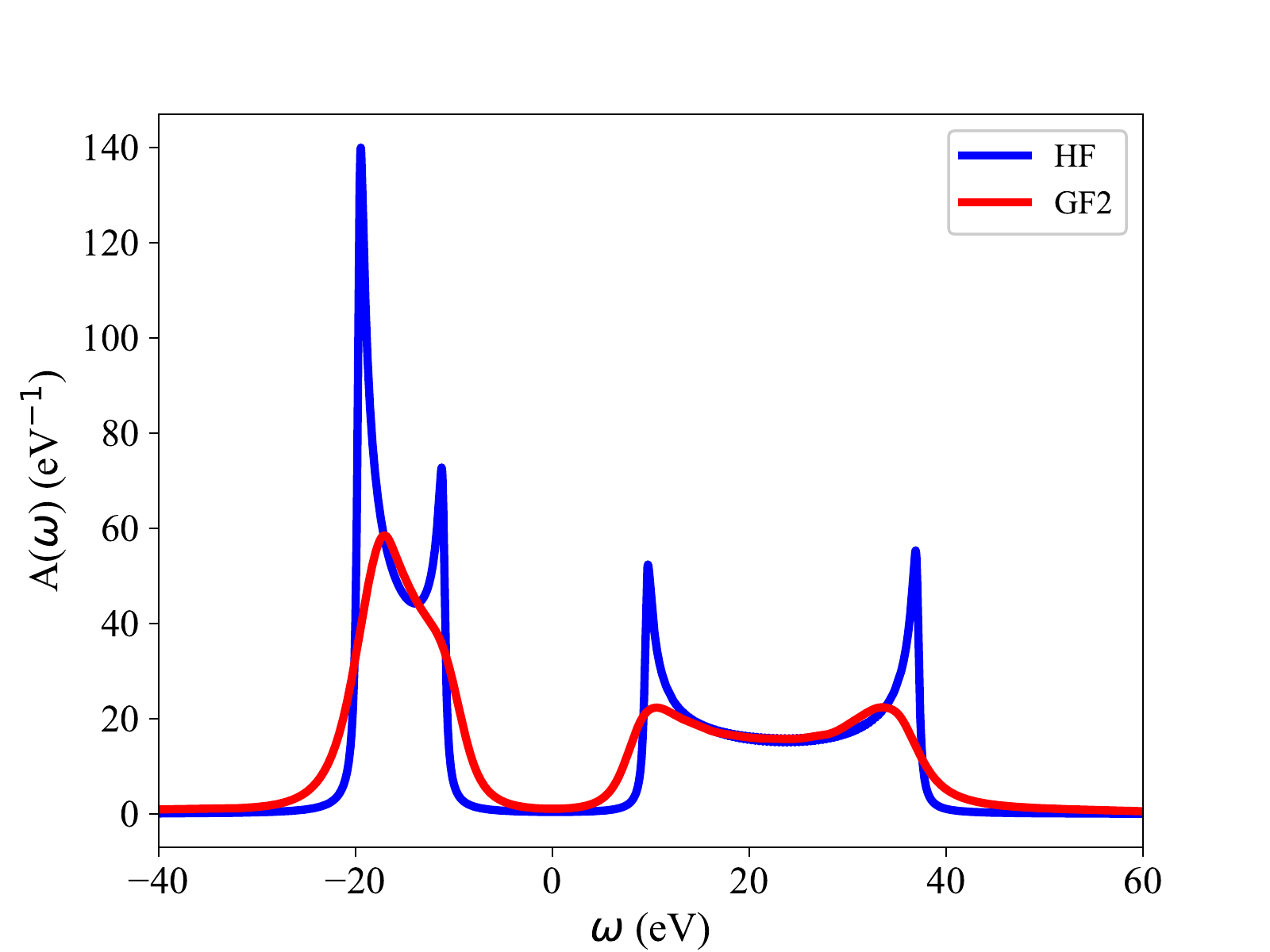} }   
   \subfloat[H$_{300}$]{\includegraphics[width=3.5in]{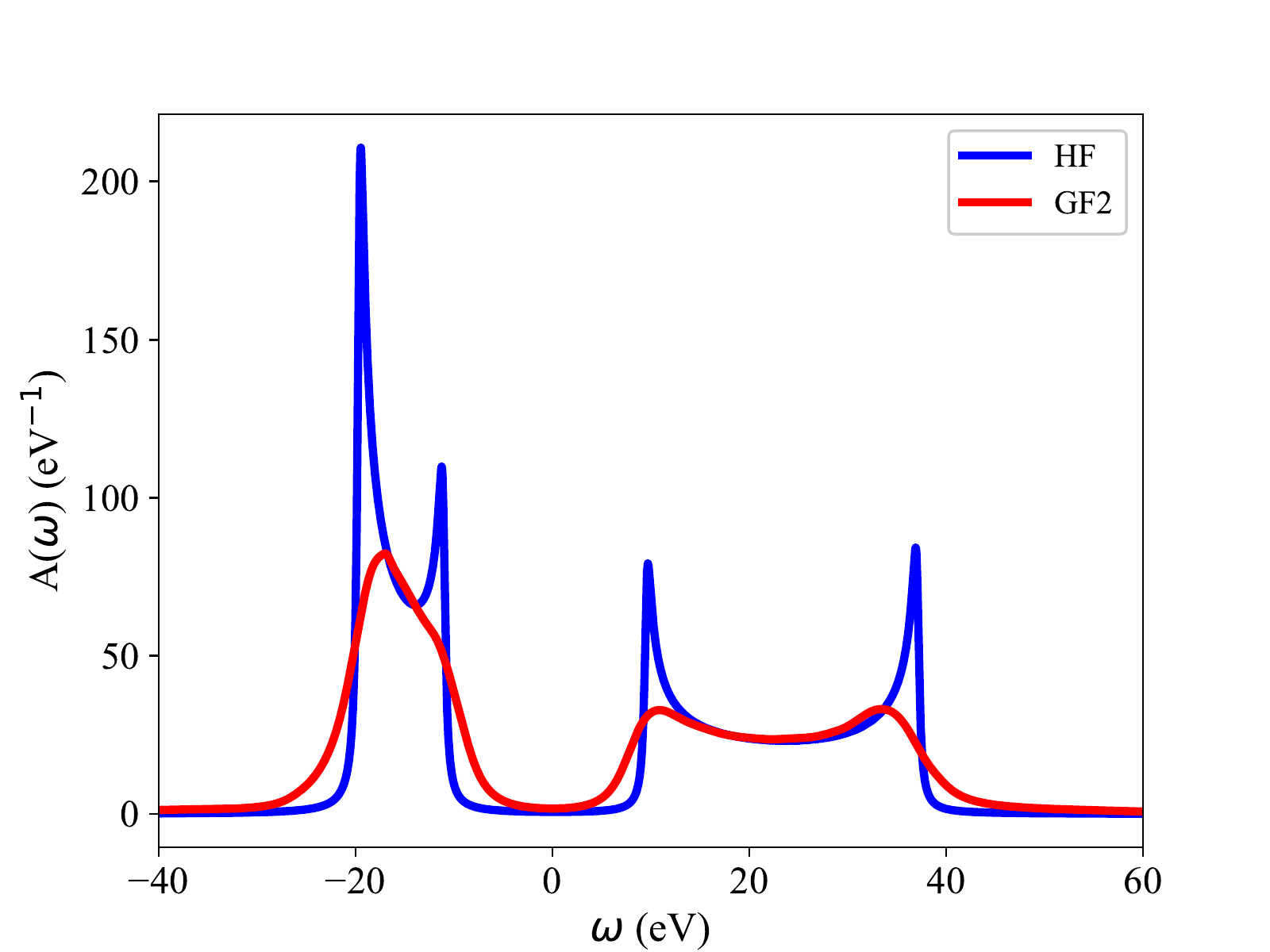} }
   \caption{Quasi-particle spectrum for H$_{20}$, H$_{80}$, H$_{200}$
     and H$_{300}$ using sto-3g basis within the HF (blue curve) and
     GF2 (red curve) approaches.}
   \label{fig:spe}
\end{figure}

In Ref.~\citen{takeshita2019stochastic}, we have reported on ground
state correlation energies for hydrogen dimer chains using stochastic
resolution of identity Matsubara second-order Green's function theory.
The same setup will be used in the current work for the real-frequency
properties. In short, we set the H-H bond distance to $\SI{0.74}
{\angstrom}$ and the long distance to $\SI{2.0} {\angstrom}$.  Minimal
basis sto-3g is used to represent GFs and the cc-pvdz-jkfit and
cc-pvdz-ri basis sets were used for the Hartree-Fock matrix fitting
and for the $4$-index ERI fitting, respectively.  The time step used
to propagate the mixed-time GF is $0.2 E_h^{-1}$.

In Fig.~\ref{fig:spe}, we plot quasi-particle spectrum for a set of
hydrogen dimer chains: H$_{20}$, H$_{80}$, H$_{200}$ and
H$_{300}$. For completeness, we also plot the spectrum from HF theory. 
For small chain length we
observe individual transition for both the valance and conduction
bands. As the length of the chain increase these feature are washed
out (more so for GF2, which contains an imaginary portion to the
self-energy which broadens the transitions) and finally a
semi-continuous density of states is formed. We also find that the
fundamental gap (quasiparticle gap) from GF2 is smaller than HF due
to electronic correlations.

\begin{figure}[htbp] 
   \centering
   \includegraphics[width=4in]{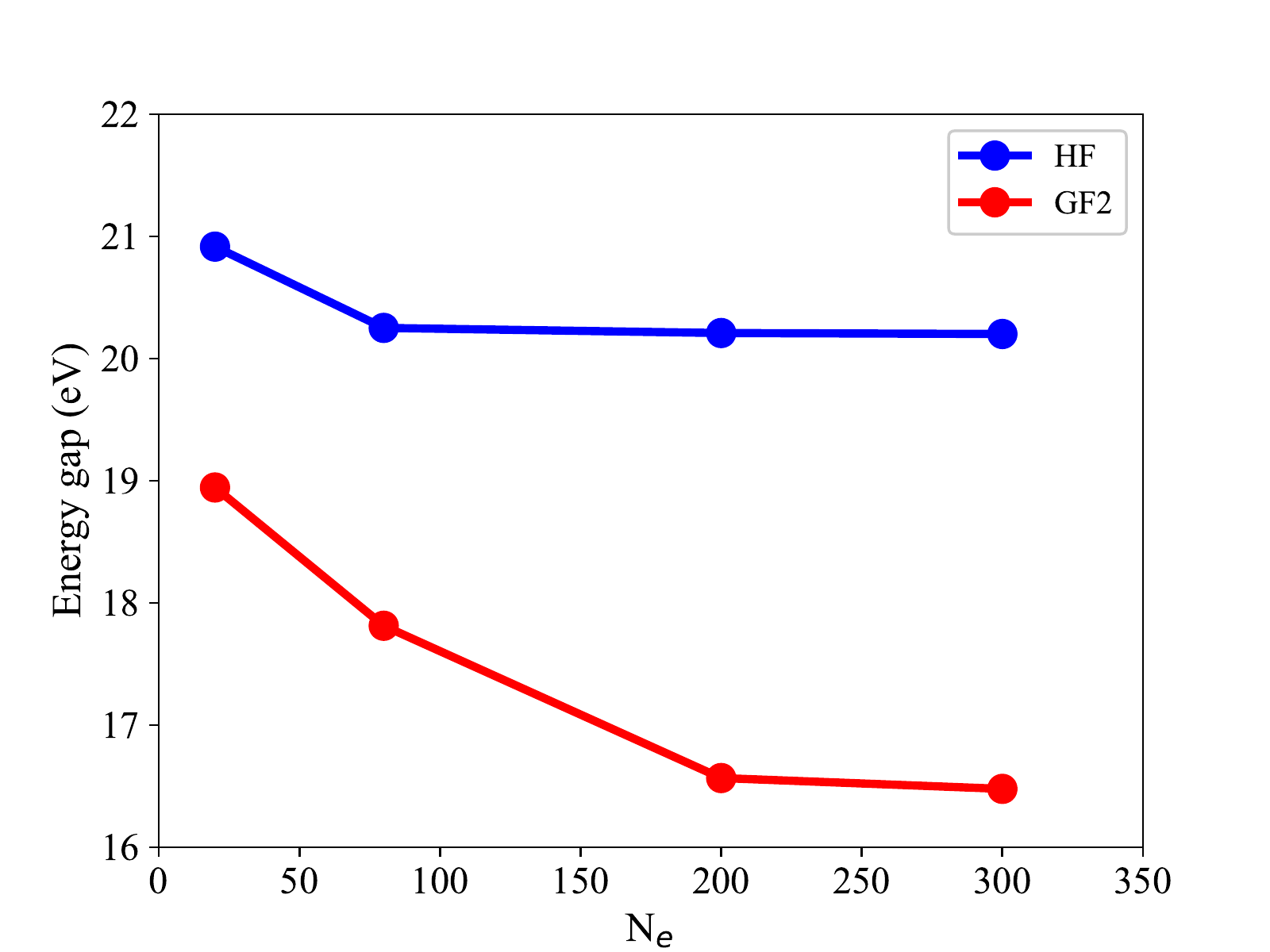} 
   \caption{Fundamental energy gaps as a function of number of
     electrons in the hydrogen dimer chain. Due to incorporation of
     electron correlations, GF2 predicts smaller fundamental energy
     gaps as compared to HF. In both cases, energy gaps decrease with
     the number of Hydrogen atoms and converge to a fixed number. }
   \label{fig:gap}
\end{figure}

In Fig. ~\ref{fig:gap}, we further plot fundamental energy gaps as a
function of the length of hydrogen dimer chain from HF and GF2. The
fundamental energy gaps are taken as the energy difference between
-IPs and EAs.  We define the IP/EA as the frequency position at half
the height of peak near HOMO/LUMO.  As mentioned above, by
incorporating electronic correlations, GF2 predicts smaller energy
gaps than HF.  Note that fundamental energy gaps decrease with the
length of Hydrogen dimer chain for both HF and GF2. Note also that
fundamental energy gaps from HF converge more quicker than GF2 as a
function of number of particles.

\begin{figure}[htbp] 
   \centering
   \includegraphics[width=4in]{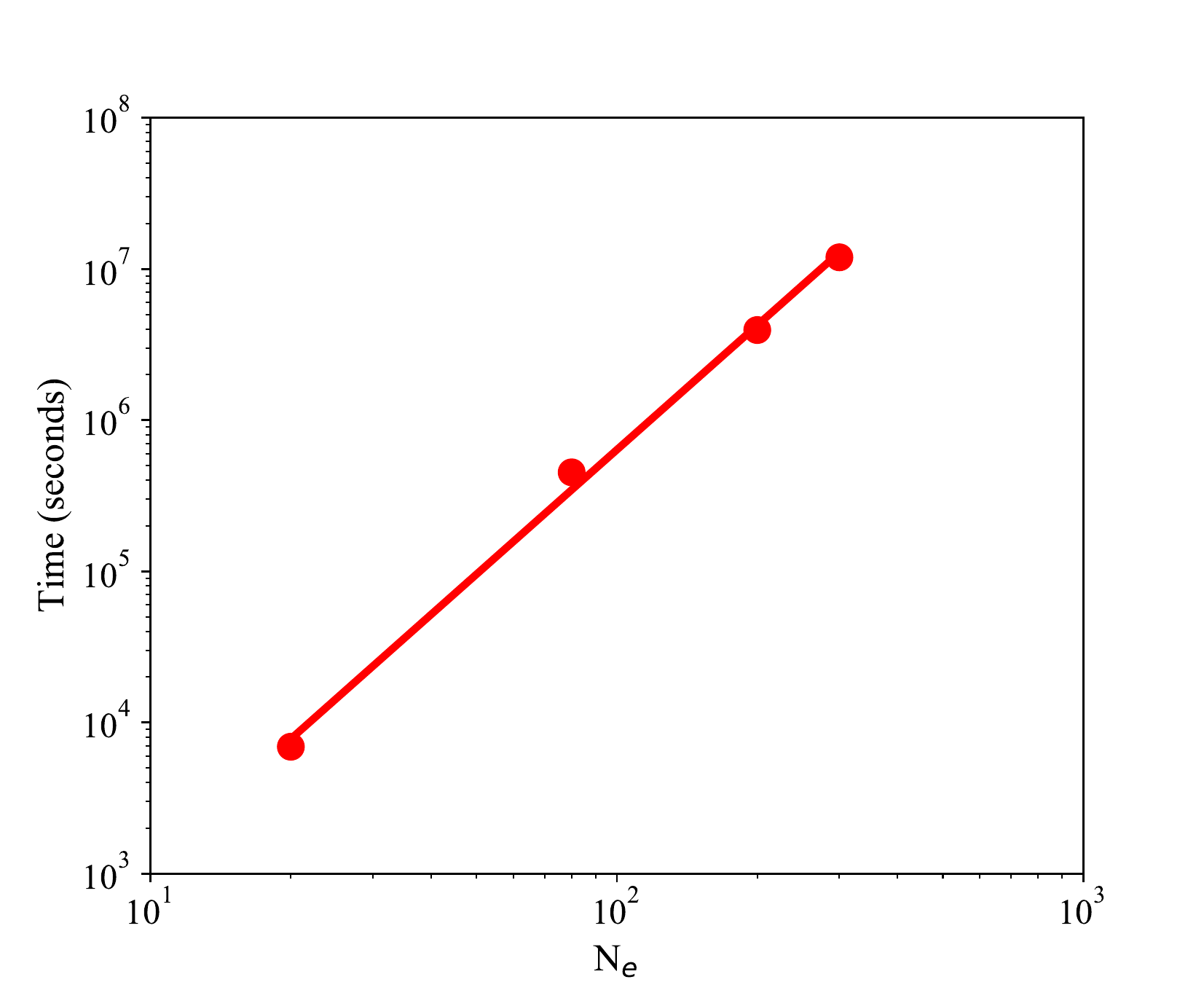} 
   \caption{Computational wall time for hydrogen dimer chains
     $N_e$ is the number of electrons. The straight line is
     a power-law fit to the date, suggesting that the scaling is
     $O(N_e^{2.7})$, slightly better than the theoretical limit of
     $O(N_e^3)$.}
   \label{fig:timing}
\end{figure}

The computational time and overall scaling of the real-time sRI-GF2
approach is summarized In Fig.~\ref{fig:timing}.  We plot the
computational wall time as a function of number of hydrogen atoms in
the chain, $N$. These results were generated with $2000$ stochastic
orbitals to reduce the error to $0.02$ eV.  We propagate the real-time
GFs to $t_{max}=200 E_h^{-1}$, which is sufficient to converge the
self-energy to within $10^{-6}$ eV of it maximal value. As mentioned
above, the computational bottleneck for real-time GF propagation is
the evaluation of the self-energy in Eq.~\ref{eq:Sigceil}, which
scales as $O(N_e^5)$. The formal scaling using the stochastic RI is
$O(N_e^3)$ (see Eqs.~(\ref{eq:sRI-SE})-(\ref{eq:sRI-SE2})). The
results for the hydrogen chain show that in practice the real-time
sRI-GF2 scales as $O(N_e^{2.7})$ on multiple $32$-core Intel-Xeon CPU
E5-2698 v3 at $2.3$GHz nodes.

\section{Conclusions} \label{sec:con}
We have developed a stochastic resolution of identity approach to
describe real-time/real-frequency spectral functions of extended
systems within the second-order Green's function formalism.  The real
time approach provides a platform to compute the ionization potentials
and electron affinities for open as well as periodic boundary
conditions. Such an approach can also be used to generate the
full-frequency quasi-particle spectral function at the same
computational cost.  The advantage of the stochastic formalism is that
it reduces the computational scaling of the real-time sRI-GF2 from
$O(N_e^5)$ to $O(N_e^3)$, as illustrated for a chain of hydrogen
dimers. This reduced scaling opens the door to study quasi-particle
excitations in extended systems within the framework of second-order
Green's function.

To access the approach, we benchmarked our real-time sRI-GF2 scheme
against a many-body perturbation technique within the GW approximation
as well as compared the calculated ionization potentials to
experimental results.  We find that the sRI-GF2 results are comparable
to the state-of-the-art self-consistent GW approach for a set of atoms
and small molecules. While GF2 lacks the sort of screening present in
the GW approximation, GF2 does include exchange effects in the
self-energy, which turn out to be significant in describing the
quasi-particle spectrum of molecules.

\begin{acknowledgement}
We would like to thank Felipe H. da Jornada and Steven G. Louie for
helpful discussion.  RB gratefully acknowledges support from the
Israel Science Foundation, grant No. 800/19. DN and ER are grateful
for support by the Center for Computational Study of Excited State
Phenomena in Energy Materials (C2SEPEM) at the Lawrence Berkeley
National Laboratory, which is funded by the U.S. Department of Energy,
Office of Science, Basic energy Sciences, Materials Sciences and
Engineering Division under contract No. DEAC02-05CH11231 as part of
the Computational Materials Sciences Program. Resources of the
National Energy Research Scientific Computing Center (NERSC), a
U.S. Department of Energy Office of Science User Facility operated
under Contract No. DE-AC02-05CH11231 are greatly acknowledged.
\end{acknowledgement}  

\providecommand{\latin}[1]{#1}
\makeatletter
\providecommand{\doi}
  {\begingroup\let\do\@makeother\dospecials
  \catcode`\{=1 \catcode`\}=2 \doi@aux}
\providecommand{\doi@aux}[1]{\endgroup\texttt{#1}}
\makeatother
\providecommand*\mcitethebibliography{\thebibliography}
\csname @ifundefined\endcsname{endmcitethebibliography}
  {\let\endmcitethebibliography\endthebibliography}{}


\begin{mcitethebibliography}{54}
\bibitem[not()]{noteForFFT}
We have set $t_{ce} = 100 E_h^{-1}$ in our calculations.\relax
\mciteBstWouldAddEndPunctfalse
\mciteSetBstMidEndSepPunct{\mcitedefaultmidpunct}
{}{\mcitedefaultseppunct}\relax
\EndOfBibitem
\end{mcitethebibliography}
\end{document}